\documentclass[twocolumn,showpacs,preprintnumbers,amsmath,amssymb,superscriptaddress]{revtex4}
\usepackage{amsmath} 
\usepackage{amssymb}
\usepackage{dcolumn}
\usepackage{graphicx}        
\def\Journal#1#2#3#4{{#1} {\bf #2}, #3 (#4)}

\def\NIMA{{ Nucl. Instrum. Methods} A}

\def\PLB{{ Phys. \mbox{Lett.}}  B}
\def\PRL{ Phys. Rev. \mbox{Lett.}}
\def\PRD{{ Phys. Rev.} D}

\def\EUPJ{{ Eur. Phys. J.} C} 
 
\def\PPNP{ Prog. Part. Nucl. Phys.} 

\newcommand{\etal}{{\it et al.}}

\begin{document}
\title{Single-spin azimuthal asymmetry in exclusive
electroproduction of $\pi^+$ mesons
}

\def\groupalberta{\affiliation{Department of Physics, University of Alberta, Edmonton, Alberta T6G 2J1, Canada}}
\def\groupargonne{\affiliation{Physics Division, Argonne National Laboratory, Argonne, Illinois 60439-4843, USA}}
\def\groupbari{\affiliation{Istituto Nazionale di Fisica Nucleare, Sezione di Bari, 70124 Bari, Italy}}
\def\groupcaltech{\affiliation{W.K. Kellogg Radiation Laboratory, California Institute of Technology, Pasadena, California 91125, USA}}
\def\groupcolorado{\affiliation{Nuclear Physics Laboratory, University of Colorado, Boulder, Colorado 80309-0446, USA}}
\def\groupdesy{\affiliation{DESY, Deutsches Elektronen-Synchrotron, 22603 Hamburg, Germany}}
\def\groupzeuthen{\affiliation{DESY Zeuthen, 15738 Zeuthen, Germany}}
\def\groupdubna{\affiliation{Joint Institute for Nuclear Research, 141980 Dubna, Russia}}
\def\grouperlangen{\affiliation{Physikalisches Institut, Universit\"at Erlangen-N\"urnberg, 91058 Erlangen, Germany}}
\def\groupferrara{\affiliation{Istituto Nazionale di Fisica Nucleare, Sezione di Ferrara and Dipartimento di Fisica, Universit\`a di Ferrara, 44100 Ferrara, Italy}}
\def\groupfrascati{\affiliation{Istituto Nazionale di Fisica Nucleare, Laboratori Nazionali di Frascati, 00044 Frascati, Italy}}
\def\groupfreiburg{\affiliation{Fakult\"at f\"ur Physik, Universit\"at Freiburg, 79104 Freiburg, Germany}}
\def\groupgent{\affiliation{Department of Subatomic and Radiation Physics, University of Gent, 9000 Gent, Belgium}}
\def\groupgiessen{\affiliation{Physikalisches Institut, Universit\"at Gie{\ss}en, 35392 Gie{\ss}en, Germany}}
\def\groupillinois{\affiliation{Department of Physics, University of Illinois, Urbana, Illinois 61801, USA}}
\def\groupliverpool{\affiliation{Physics Department, University of Liverpool, Liverpool L69 7ZE, United Kingdom}}
\def\groupwisconsin{\affiliation{Department of Physics, University of Wisconsin-Madison, Madison, Wisconsin 53706, USA}}
\def\groupmit{\affiliation{Laboratory for Nuclear Science, Massachusetts Institute of Technology, Cambridge, Massachusetts 02139, USA}}
\def\groupmichigan{\affiliation{Randall Laboratory of Physics, University of Michigan, Ann Arbor, Michigan 48109-1120, USA }}
\def\groupmoscow{\affiliation{Lebedev Physical Institute, 117924 Moscow, Russia}}
\def\groupmunich{\affiliation{Sektion Physik, Universit\"at M\"unchen, 85748 Garching, Germany}}
\def\groupnewmexico{\affiliation{Department of Physics, New Mexico State University, Las Cruces, New Mexico 88003, USA}}
\def\groupnikhef{\affiliation{Nationaal Instituut voor Kernfysica en Hoge-Energiefysica (NIKHEF), 1009 DB Amsterdam, The Netherlands}}
\def\groupstpetersburg{\affiliation{Petersburg Nuclear Physics Institute, St. Petersburg, Gatchina, 188350 Russia}}
\def\groupregensburg{\affiliation{Institut f\"ur Theoretische Physik, Universit\"at Regensburg, 93040 Regensburg, Germany}}
\def\grouprome{\affiliation{Istituto Nazionale di Fisica Nucleare, Sezione Roma 1, Gruppo Sanit\`a and Physics Laboratory, Istituto Superiore di Sanit\`a, 00161 Roma, Italy}}
\def\groupsimonfraser{\affiliation{Department of Physics, Simon Fraser University, Burnaby, British Columbia V5A 1S6, Canada}}
\def\grouptriumf{\affiliation{TRIUMF, Vancouver, British Columbia V6T 2A3, Canada}}
\def\grouptokyo{\affiliation{Department of Physics, Tokyo Institute of Technology, Tokyo 152, Japan}}
\def\groupamsterdam{\affiliation{Department of Physics and Astronomy, Vrije Universiteit, 1081 HV Amsterdam, The Netherlands}}
\def\groupyerevan{\affiliation{Yerevan Physics Institute, 375036, Yerevan, Armenia}}


\groupalberta
\groupargonne
\groupbari
\groupcaltech
\groupcolorado
\groupdesy
\groupzeuthen
\groupdubna
\grouperlangen
\groupferrara
\groupfrascati
\groupfreiburg
\groupgent
\groupgiessen
\groupillinois
\groupliverpool
\groupwisconsin
\groupmit
\groupmichigan
\groupmoscow
\groupmunich
\groupnewmexico
\groupnikhef
\groupstpetersburg
\groupregensburg
\grouprome
\groupsimonfraser
\grouptriumf
\grouptokyo
\groupamsterdam
\groupyerevan


\author{A.~Airapetian}  \groupyerevan
\author{N.~Akopov}  \groupyerevan
\author{Z.~Akopov}  \groupyerevan
\author{M.~Amarian}  \grouprome \groupyerevan
\author{E.C.~Aschenauer}  \groupzeuthen
\author{H.~Avakian}  \groupfrascati
\author{R.~Avakian}  \groupyerevan
\author{A.~Avetissian}  \groupyerevan
\author{E.~Avetissian}  \groupyerevan
\author{P.~Bailey}  \groupillinois
% ...

\collaboration{The HERMES Collaboration} \noaffiliation

\date{\today}

\begin{abstract}
A single-spin asymmetry in the distribution
of exclusively produced $\pi^+$ mesons azimuthally around the virtual
photon direction relative to the lepton scattering plane
has been measured for the first time 
in deep-inelastic scattering of positrons off longitudinally polarized protons.
Integrated over the experimental acceptance, the $\sin \phi$ moment of the polarization
asymmetry of the cross section is measured to be
$-0.18 \pm 0.05\,\mbox{(stat.)} \pm 0.02\,\mbox{(syst.)}$. 
The asymmetry is also studied as a function of the relevant kinematic 
variables,
and its magnitude is found to grow 
with decreasing $x$ and increasing $-t$ and vanish at $t \rightarrow t_{min}$ (where $x$ is
the \mbox{Bjorken} scaling variable and $t$ is the squared 
four-momentum transferred to the nucleon).
\end{abstract}


\maketitle
 
\begin{figure*}[!t]
\includegraphics[width=15.0cm]{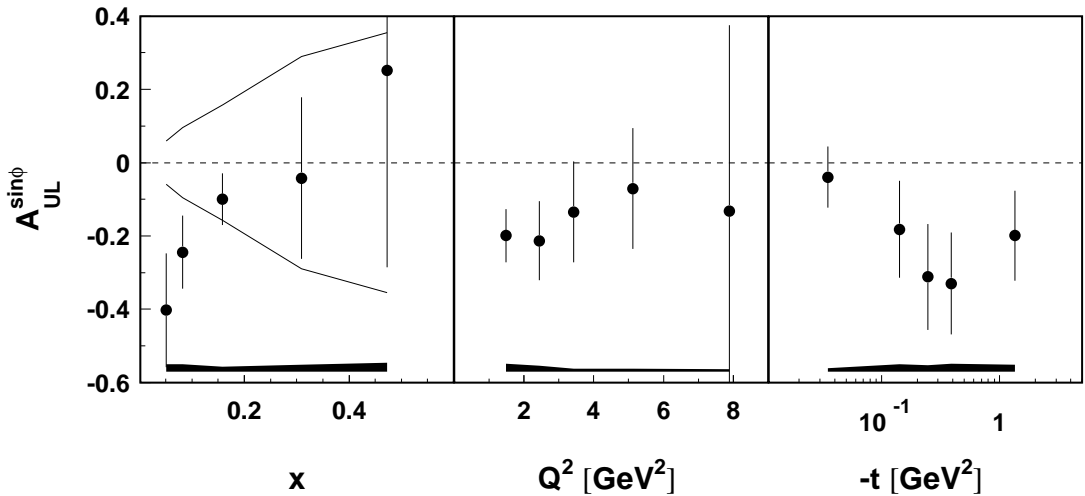}
\caption{Kinematic dependence of $ A_{\mathrm{UL}}^{\sin \phi}$ on the variables
$x$, $Q^2$, and $t$ for the reaction $e^+ + \vec{p} \rightarrow e^{\prime +} + n  + \pi^+$. 
The error bars and bands represent the statistical and
systematic uncertainties, respectively. The solid lines show the upper limits for any
asymmetry arising from the transverse target polarization component.}
\label{aulall}
\end{figure*}

The interest in hard exclusive processes has grown since
a QCD factorization theorem was
proved for the hard exclusive production
of mesons by longitudinal virtual photons (helicity $0$) \cite{cfs}.
The Generalized Parton Distribution functions (GPDs) \cite{Mul94,Rad97}
appearing in this factorization scheme
are of great interest because they
can be related to parton angular momentum distribution functions
that are not directly constrained or inaccessible through semi-inclusive or inclusive
measurements \cite{xiji,Goeke01}.
\phantom{\ref{aulall}}

While unpolarized GPDs (typically designated as $E$ and $H$) can be probed through
exclusive vector meson production,
 polarized GPDs (typically designated as $\tilde{E}$ and $\tilde{H}$) can be probed through exclusive pseudoscalar meson
production without the need for a polarized target or beam. 
However, only a quadratic
combination of GPDs appears in the unpolarized cross section for exclusive meson electroproduction,
and so several independent observables are needed to disentangle the various distributions
\cite{diehl_erhic}. 
Additional observables
can be accessed through the measurement of the polarized cross section.
For example, it has been predicted \cite{fpsv} that for 
the exclusive production of $\pi^+$ mesons 
from a transversely polarized target by longitudinal
virtual photons,
the interference between the 
pseudoscalar ($\tilde{E}$) and pseudovector ($\tilde{H}$) amplitudes leads
to a large target-related single-spin asymmetry in the distribution of the angle $\phi$.
Here $\phi$ is the azimuthal angle of the pion around the virtual photon momentum
relative to the lepton scattering plane.
The predicted asymmetry is of order unity.
Moreover, the scaling region of this asymmetry (where
corrections proportional to powers of $1/Q$ are small) is reached 
at lower $Q^2$ of the virtual photon than for the absolute cross section \cite{fpps}.
It has also been shown that corrections that are next-to-leading order (NLO) in $\alpha_s$ cancel in the transverse
asymmetry \cite{Belitsky_Muller}.
No factorization theorem has been proved for transverse virtual
photons (helicity $\pm 1$) , but their
contribution to the cross section is predicted to be suppressed by 
at least a power of $1/Q$ \cite{cfs}.

In the case of electroproduction from a target polarized \textit{longitudinally}
with respect to the lepton beam momentum, 
a small transverse component ($S_{\perp}$) of the target polarization 
orthogonal to the virtual photon direction
does appear,
along with the dominant longitudinal component ($S_{\parallel}$).
The polarized cross section ($\sigma_S$) for this reaction
has the form
\cite{polyakov_dis00,Diehl_ringberg}
\begin{equation}
\sigma_S \sim [S_{\perp} \sigma_{\mathcal L} + S_{\parallel} \sigma_{\mathcal L \mathcal T}]\;
		A_{\mathrm{UL}}^{\sin \phi} \, \sin \phi \;,
\label{form}
\end{equation}
where $A_{\mathrm{UL}}^{\sin \phi}$ is the $\sin \phi$ moment of the polarization asymmetry.
The subscripts $\mathrm{U}$ and $\mathrm{L}$ indicate the use of 
an unpolarized beam and longitudinally polarized target, respectively.
Equation~(\ref{form}) shows that in the case of a longitudinally polarized
target an additional term appears which contains the interference of longitudinal ($\mathcal L$) and transverse ($\mathcal T$) virtual photon amplitudes. 
Since a factorization theorem was not proved yet for transverse virtual photons
quantitative predictions for this term would
require a calculation in next-to-leading twist \cite{Belitsky_Muller}.

\begin{figure}[bt]
\centerline{\includegraphics[width=7.5cm]{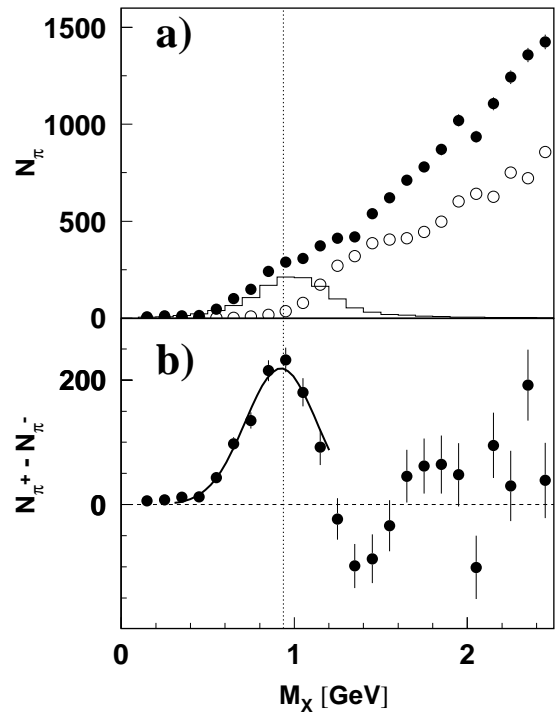}}
\caption{a) Missing mass distribution for
$\pi^+$ (filled circles) and $\pi^-$ (empty circles) electroproduction at HERMES.
The histogram is a Monte Carlo prediction for exclusive $\pi^+$ 
production with arbitrary normalization.
b) Difference between the $\pi^+$ and normalized $\pi^-$ distributions (see text).
The curve is a Gaussian fit to the data for $0.4<M_X<1.2$ GeV.
The vertical dotted 
line indicates the nucleon mass. 
The error bars represent the statistical uncertainties.
}
\label{mmfig}
\end{figure}

This letter reports on the first measurement of a single-spin asymmetry
in the exclusive reaction $e^+ + \vec{p} \rightarrow e^{\prime +} + n  + \pi^+$.
The relevant kinematic variables of this process in the target rest frame
are the space-like squared four-momentum $-Q^2$ of the exchanged virtual photon
with energy $\nu$, the squared four-momentum $t$ transferred
to the nucleon,
the \mbox{Bjorken} scaling variable $x \equiv Q^2 / 2M\nu$, where
$M$ is the proton mass, and
the azimuthal angle $\phi$ described above. The sign of $\phi$ is
given by $(\vec{e^+} \times \vec{e^{\prime +}}) \cdot \vec{\pi^+} /
         |(\vec{e^+} \times \vec{e^{\prime +}}) \cdot \vec{\pi^+}|$.
\\ \indent
The data were collected in 1997
using  a longitudinally polarized hydrogen gas target in the 27.6~GeV HERA
positron storage ring at DESY.
While the lepton beam was usually polarized oppositely to its direction,
a small data sample taken with parallel beam polarization has shown that
the measured asymmetry is independent of the lepton
beam polarization.
The average target polarization was 0.88 with a fractional
uncertainty of 5$\%$ \cite{g1p}.
The scattered positron and  the produced hadron
were detected by the HERMES spectrometer {\cite{hermes:spectrometer}}.
Positrons were distinguished from hadrons with an average efficiency of
99$\%$ and a hadron contamination of less than 1$\%$ using the information
from an electromagnetic calorimeter, a transition radiation detector,
a preshower scintillator detector, and a threshold \v{C}erenkov  detector.
The kinematic requirements imposed on the scattered positrons were
$Q^2 > 1$~GeV$^2$, $0.02 < x < 0.8$ and
an invariant mass squared of the initial photon-nucleon system $W^2 > 4$~GeV$^2$.
\\ \indent      
The recoiling neutron was not detected, and so exclusive production
of mesons was
selected by requiring that the missing mass ($M_X$) of the reaction
$e^+ + p \rightarrow e^{\prime +} + \pi^+ + X$ corresponded to the nucleon mass.
The missing mass distribution for $\pi^+$
is shown in Fig.~\ref{mmfig}a (filled circles), 
for the pion momentum range from 4.9 to 14.0 GeV.
These cuts on the pion's momentum serve only to restrict
to a kinematic region
in which the \v{C}erenkov detector provides pion identification,
and were not used in any other part of the analysis.
A large number of events is observed in the missing mass
region around the nucleon mass. However, from the $\pi^+$ missing
mass distribution alone, it is not possible to separate the
exclusive channel $e^+ + p \rightarrow e^{\prime +} + \pi^+ + n$ from
the neighboring (defined as non-exclusive) channels $\pi^+ + \Delta^0$,
$\pi^+ + (N \pi)$ and
$\pi^+ + (N \pi \pi)$, which can be
smeared into the $\pi^+ + n$ region due to the limited experimental
resolution.
The histogram which is represented by a solid line in Fig.~\ref{mmfig}a is an arbitrarily normalized
Monte Carlo simulation of the detector response
to exclusive $\pi^+$ generated according to
the predictions of refs.~\cite{piller,vgg} for the HERMES kinematics.
It yields an experimental $M_X$ resolution for the exclusive $\pi^+$ channel of about 230 MeV,
which is larger than the separation from the
non-exclusive channels mentioned before. The latter ones are also present
in the $e^+ + p \rightarrow e^{\prime +} + \pi^- + X$ process and are shown in Fig.~\ref{mmfig}a
as empty circles,
while exclusive $\pi^-$ production with a nucleon in the final state is forbidden
by charge conservation. As for $\pi^+$, also non-exclusive $\pi^-$ events are smeared into the 
exclusive region.
Therefore, the non-exclusive $\pi^+$ background
was estimated from the normalized number of $\pi^-$ passing
the same cuts as the $\pi^+$ 
and found in the same exclusive missing mass region.

In Fig.~\ref{mmfig}b, the difference between the $\pi^+$ and $\pi^-$ missing mass distributions
is shown, after normalizing the integral of the $\pi^-$ distribution with respect to the
$\pi^+$ one in the range  $1.3 < M_X < 2.0$ GeV.
This interval sees strong contributions from non-exclusive background from both 
resonant ($\pi^+ + \Delta^0$) and non-resonant 
($\pi^+ + (N\pi)$ and $\pi^+ + (N\pi \pi$)) channels.
The resulting background normalization factor is about $1.6$, with a systematic 
uncertainty of $13\%$.
The systematic uncertainty was estimated by using different normalization
regions i.e. $1.2<M_X<1.5$ GeV and $2.0<M_X<2.5$ GeV.
The difference
between the $\pi^+$ and $\pi^-$ missing mass distributions
shows a clear peak centered at the nucleon mass. 
The curve  in Fig.~\ref{mmfig}b is a Gaussian
fit to the data.
The deficit seen in the distribution 
at $M_X \approx 1.2-1.6$ GeV is due to a difference in the 
relative contribution of the resonant and non-resonant  channels
to the  $\pi^+$ and $\pi^-$ yields.

The $\phi$ dependence of the polarized cross section
appears in
the cross section asymmetry for exclusively produced $\pi^+$, which is defined by
\begin{equation}
A(\phi) = \frac{1}{|S|} \,\frac{N_e^{\uparrow}(\phi) - N_e^{\downarrow}(\phi)}{N_e^{\uparrow}(\phi) + N_e^{\downarrow}(\phi)}.
\label{csa_e}
\end{equation}
Here $N_e$ represents the yield of exclusive $\pi^+$, the superscript
$\uparrow(\downarrow)$ denotes a target polarization direction 
anti-parallel (parallel) to the positron beam momentum,
and $S$ is the degree of polarization of the target protons.
All the tracks identified as hadrons were assumed to be pions,
as the requirement that the missing mass corresponds to the nucleon mass
removes all the hadrons heavier than the pion from the data sample. 
An analysis requiring pion identification by the \v{C}erenkov
detector was performed and gave similar results but with reduced statistics.

\begin{figure}[bt]
\centerline{\includegraphics[width=7.0cm]{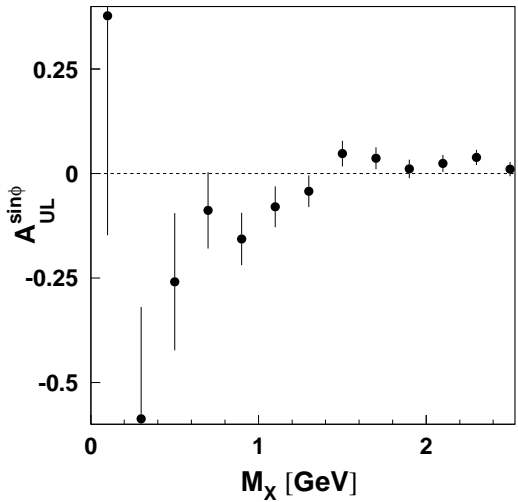}}
\caption{$A_{\mathrm{UL}}^{\sin \phi}$ for the
$e^+ + \vec{p} \rightarrow e^{\prime +} +  h^+ + X$ reaction as a function of
missing mass.
The error bars represent the statistical uncertainties.}
\label{semih}
\end{figure}

To evaluate the asymmetry defined by equation (\ref{csa_e}),
a background correction has been applied.
The total number of events $N(\phi)$ in each bin is the sum of 
exclusive events $N_e(\phi)$ and background events $N_{bg}(\phi)$.
Equation (\ref{csa_e}) can thus be written
\begin{equation}
A(\phi) =\frac{1}{|S|} \, \frac{N^{\uparrow}(\phi) - N^{\downarrow}(\phi) \; - |S| A_{bg}(\phi) N_{bg}(\phi)}
                      {N^{\uparrow}(\phi) + N^{\downarrow}(\phi)  - N_{bg}(\phi)}
\label{csa_bg}
\end{equation} 
where $A_{bg}(\phi) = \frac{1}{|S|} \,\frac{N_{bg}^{\uparrow}(\phi) - N_{bg}^{\downarrow}(\phi)}{N_{bg}^{\uparrow}(\phi) + N_{bg}^{\downarrow}(\phi)}$ is the asymmetry of the background. As shown in equation (\ref{csa_bg}), 
only two parameters are needed to correct for background
processes: the background yield, $N_{bg} (\phi) = N_{bg}^{\uparrow}(\phi) + N_{bg}^{\downarrow}(\phi)$,
and its asymmetry, $A_{bg} (\phi)$. As discussed earlier, the background yield in each bin
was estimated from the normalized number of $\pi^-$ events passing all the cuts applied to
 $\pi^+$ events. Since the background originates
from the smearing of events occurring at higher missing mass, 
the background asymmetry was estimated
to be the asymmetry from the neighboring 
missing mass region where the contribution of exclusive $\pi^+$ events is negligible.
The $\sin \phi$ moment of the uncorrected asymmetry ($A(\phi)$ in equation (\ref{csa_bg})
with $N_{bg} = 0$) is shown in Fig.~\ref{semih} as a function of $M_X$. All the other moments were found to be compatible with zero.
In the missing mass region from 1.3 to 2.0 GeV, $A_{\mathrm{UL}}^{\sin \phi}$
ranges between $-0.05$ and $+0.05$ with an average value of $0.019 \pm 0.014$.
This asymmetry is compatible with that measured for semi-inclusive $\pi^+$
production \cite{HERMES:ssa:pi+}.
Therefore the $\sin \phi$ moment of the background asymmetry has been taken conservatively 
to be $0 \pm 0.1$.
It is important
to stress that the exclusive asymmetry computed by means of equation 
(\ref{csa_bg}) (i.e. after background correction)
is independent of the missing mass cut within the range
$0.7< M_X<1.5 \;\mbox{GeV}$, despite the fact that the signal-to-background
ratio ($N_e/N_{bg}$) goes from 20 to 0.8 in the same interval. 
However, a missing mass cut has been applied to optimize the statistical accuracy.
For a cut at high missing mass the
statistical error on $A(\phi)$ given by equation (\ref{csa_bg})
is dominated by the statistics of the background sample.
For a cut at low missing mass it is determined by the (reduced) number of exclusive events.
The best statistical accuracy is obtained by requiring $M_X<1.05 \; \mbox{GeV}$.
For this cut, the inferred signal to background ratio is about 6 and therefore, 
the result depends only weakly on any systematic uncertainty of the background yield or
its  asymmetry.
All the results presented in this paper were obtained by requiring $M_X<1.05 \; \mbox{GeV}$.

\begin{figure}[bt]
\centerline{\includegraphics[width=7.0cm]{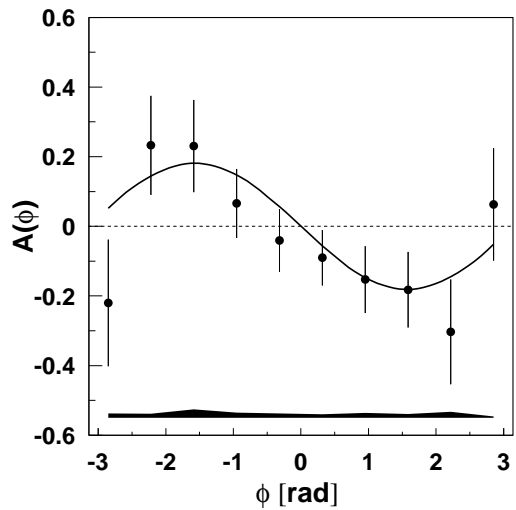}}
\caption{Cross section asymmetry $A(\phi)$ averaged over $x$, $Q^2$, and $t$ for the 
reaction $e^+ + \vec{p} \rightarrow e^{\prime +} + n  + \pi^+$.
The curve is the best fit to the data by
$A(\phi) = A_{\mathrm{UL}}^{\sin \phi} \cdot \sin \phi$
with $ A_{\mathrm{UL}}^{\sin \phi} = -0.18 \pm 0.05$ at a reduced $\chi^2$ of 0.8.
The error bars and bands represent the statistical and
systematic uncertainties, respectively.}
\label{aulphi}
\end{figure}

The cross section asymmetry integrated over $x$, $Q^2$ and $t$ is shown in Fig.~\ref{aulphi}. 
The average values of the kinematic variables are
$\langle x \rangle = 0.15$, $\langle Q^2 \rangle=2.2\;\mbox{GeV}^2$ and 
$\langle t \rangle = -0.46\;\mbox{GeV}^2$ (with $75\%$ of the events occurring at $|t|<0.5\;\mbox{GeV}^2$).
The data show a large asymmetry in the distribution versus azimuthal angle $\phi$, with 
a clear $\sin \phi$ dependence.
A fit to this dependence of the form
\begin{equation}
A(\phi) = A_{\mathrm{UL}}^{\sin \phi} \cdot \sin \phi
\end{equation}
yields $A_{\mathrm{UL}}^{\sin \phi} = -0.18 \pm 0.05 \pm 0.02$ with a reduced $\chi^2$ of 0.8.
The background correction described above and its associated uncertainty account for $-0.024 \pm 0.011$~(syst) in this value.
A fit to $A(\phi)$ using a more general function
$A(\phi) = a \sin \phi + b \sin 2 \phi + c \cos \phi + d \cos 2 \phi$
yields values for $b$, $c$, and $d$ that are compatible with zero, showing the dominant
role of the $\sin \phi$ contribution.
The results of this fit are listed in Table \ref{tablefit}.
The band in Fig.~\ref{aulphi} represents the systematic uncertainty from background yield and asymmetry,
and from target polarization.
It has been shown by a Monte Carlo simulation \cite{Tho01} of exclusive events
that the momentum resolution
of the HERMES spectrometer does not affect the magnitude of the measured
asymmetry over the kinematic range of the measurement, and that acceptance
effects cancel in equation (\ref{csa_e}).

Current theoretical calculations \cite{fpsv,fpps,Belitsky_Muller}
for a transversely polarized target are in agreement with each other and 
predict a large asymmetry. 
The transverse component of the target polarization in the virtual photon frame is given by 
\begin{equation}
\label{sperp}
  S_{\perp}=|S| \sin{\theta}_{\gamma}=|S|
\sqrt{\frac{4M^2x^2}{Q^2+4M^2x^2}
    (1-y-\frac{M^2x^2y^2}{Q^2})},
\end{equation}
where $\theta_{\gamma}$ is the virtual photon emission angle, $E$ is the lepton beam
energy and $y=\nu/E$  is the fraction of the lepton's energy 
carried off by the virtual photon.
The average value
of  $\sin \theta_{\gamma}$ for the
present measurement  is $0.16$.
As discussed before, estimates for the asymmetry arising from
the longitudinal component of the target polarization are not yet available.

The dependence of $ A_{\mathrm{UL}}^{\sin \phi}$ on the individual kinematic variables
$x$, $Q^2$, and $t$ is shown in Fig.~\ref{aulall} and reported in Table \ref{tabledata}.
The table shows that these three kinematic quantities are
strongly correlated by the experimental conditions.
The absolute magnitude of the asymmetry shows a clear rise 
with decreasing $x$. The values of the relative transverse target polarization
component ($\langle \sin \theta_{\gamma} \rangle$) are given in Table \ref{tabledata} and shown with both signs
by the two full lines in the first panel of Fig.~\ref{aulall}, indicating the
limits for any asymmetry arising from that component. 
At low $x$ the asymmetry must arise
from the longitudinal target polarization component, independent of any model. 
No strong $Q^2$ dependence is observed within the statistical accuracy.
Finally, the asymmetry is seen to increase in absolute magnitude with $-t$ and vanish in the forward limit ($t \rightarrow t_{min}$), where the pion
momentum becomes collinear to the virtual photon momentum.
\begin{table}[tb]
\begin{tabular}{|c|c|c|c|}
\hline
a	&b 	&c	&d \\
\hline
$-0.19 \pm 0.06$	&$0.05 \pm 0.05$	&$-0.03 \pm 0.05$	&$-0.03 \pm 0.05$ \\
\hline
\end{tabular}
\caption{Results of the fit of $A(\phi)$ (Fig.~\ref{aulphi}) to the form
$A(\phi) = a \sin \phi + b \sin 2 \phi + c \cos \phi + d \cos 2 \phi$. The reduced 
$\chi^2$ is 0.9.}
\label{tablefit}
\end{table}
\begin{table}[tb]
\begin{tabular}{|c|c|c|c|c|}
\hline
        &$\langle x \rangle$ &$\langle Q^2 \rangle$     &$\langle \sin \theta_{\gamma} \rangle$ &$ A_{\mathrm{UL}}^{\sin \phi}$  \\
        & & $[\mbox{GeV}^2]$    & & \\
\hline
$x$ &&&&\\
0.05    & &1.3 &0.06 & $-0.40      \pm 0.16      \pm 0.02$ \\
0.08    & &1.6 &0.10 & $-0.24      \pm 0.10      \pm 0.02$ \\
0.16    & &2.6 &0.16 & $-0.10      \pm 0.07      \pm 0.01$ \\
0.31    & &3.6 &0.29 & $-0.04      \pm 0.22      \pm 0.02$ \\
0.47    & &5.0 &0.36 & $\phantom{-}0.25      \pm 0.54      \pm 0.02$   \\
\hline
\hline
$Q^2$   &&&&  \\
$[\mbox{GeV}^2]$ &&&& \\
1.5  &0.12 &&0.15 &$-0.20     \pm 0.07   \pm   0.02$ \\
2.4  &0.17 &&0.17 &$-0.21     \pm 0.11   \pm   0.02$ \\
3.4  &0.21 &&0.17 &$-0.13     \pm 0.14   \pm   0.01$ \\
5.1  &0.26 &&0.16 &$-0.07     \pm 0.17   \pm   0.01$ \\
7.9  &0.38 &&0.19 &$-0.13     \pm 0.51   \pm   0.01$ \\
\hline
\hline
$-t$    &&&&  \\
$[\mbox{GeV}^2]$&&&& \\
0.04    &0.11   &2.2    &0.11 &$-0.04 \pm 0.08 \pm 0.01$        \\
0.14    &0.13   &2.4    &0.13 &$-0.18 \pm 0.13 \pm 0.02$        \\
0.25    &0.14   &2.3    &0.14 &$-0.31 \pm 0.15 \pm 0.02$        \\
0.39    &0.16   &2.5    &0.16 &$-0.33 \pm 0.14 \pm 0.02$        \\
1.34    &0.24   &2.8    &0.24 &$-0.20 \pm 0.12 \pm 0.02$        \\
\hline
\end{tabular}
\caption{$A_{\mathrm{UL}}^{\sin \phi}$ as a function of $x$, $Q^2$, and $t$.}
\label{tabledata}
\end{table} 

In summary, a single-spin azimuthal asymmetry has been measured
for the first time
in the exclusive electroproduction of $\pi^+$ mesons from a longitudinally polarized
proton target.
The measured asymmetry is  large and negative. The dependence 
of this asymmetry on the kinematic variables $x$, $Q^2$, and $t$ has been investigated, 
and its magnitude is found to increase at low $x$ and at large $|t|$.
Further, the data show that the longitudinal target polarization
component with respect to the virtual photon direction provides the dominant contribution to the measured asymmetry at small $x$.
A next-to-leading twist calculation would thus be required for a complete
description of the measurement.
The present data, in combination with
future HERMES measurements on a transversely
polarized target \cite{wdn} for which a large asymmetry
is expected, opens the way to experimentally disentangle
the asymmetry arising from the two target polarization components.
This information may provide 
a first glimpse of certain unknown Generalized Parton Distributions.

We thank M.~Diehl, M.~Guidal, D.~M\"uller, G.~Piller, M.V.~Polyakov, M.~Strikman, and O.~Teryaev
for many interesting discussions on this subject.
We gratefully acknowledge the DESY management for its support and
the DESY staff and the staffs of the collaborating institutions.
This work was supported by the FWO-Flanders, Belgium;
the Natural Sciences and Engineering Research Council of Canada;
the INTAS and TMR network contributions from the European Community;
the German Bundesministerium f\"ur Bildung und For\-schung;
the Deutsche Forschungsgemeinschaft (DFG);
the Deutscher Akademischer Austauschdienst(DAAD);
the Italian Istituto Nazio\-nale di Fisica Nucleare (INFN);
Monbusho International Scientific  Research Program, JSPS, and Toray
Science Foundation of Japan;
the Dutch Foundation for Fundamenteel Onderzoek der Materie (FOM);
the U.K. Particle Physics and Astronomy Research Council; and
the U.S. Department of Energy and National Science Foundation.        

\newpage

\end{document}